\documentclass[useAMS,usenatbib]{mn2e}
\usepackage{epsfig}
\title[High-redshift Galaxies]
{ The specific star formation rate of high redshift galaxies: the case for two modes of star formation}
\author[Khochfar \& Silk]
  {Sadegh Khochfar,$^{1}$\thanks{sadeghk@mpe.mpg.de}
  Joseph Silk,$^{2}$ \\
 $^1$ Max-Planck-Institute for Extraterrestrial Physics, Giessenbachstrasse 1, 85748 Garching, Germany\\
  $^2$ Department of Physics, Denys Wilkinson Building, Keble Road, Oxford OX1 3RH, United 
Kingdom\\
   }
\date{Released 2010 Xxxxx XX}

\pagerange{\pageref{firstpage}--\pageref{lastpage}} \pubyear{2010}

\def\simpropto{\lower.2ex\hbox{$\; \buildrel \propto \over \sim \;$}}
\def\ltsim{\lower.5ex\hbox{$\; \buildrel < \over \sim \;$}}
\def\gtsim{\lower.5ex\hbox{$\; \buildrel > \over \sim \;$}}

\begin{document}
\newcommand{\mnras}{MNRAS}
\newcommand{\apj}{ApJ}
\newcommand{\apjl}{ApJL}
\newcommand{\aap}{A\&A}
\newcommand{\apjs}{ApJS}
\newcommand{\pasp}{PASP}
\newcommand{\nat}{Nature}

\label{firstpage}

\maketitle

\begin{abstract}
We study the specific star formation rate (SSFR) and its evolution at $z\gtsim 4$, in models of galaxy formation, where the star formation is driven by cold  accretion flows. We show that constant star formation and feedback efficiencies cannot reproduce the observed trend of SSFR with stellar mass and its observed lack of evolution at $z>4$. Model galaxies with $\log(M_*) \ltsim 9.5$ M$_{\odot}$ show systematically lower specific star formation rates by orders of magnitudes, while massive galaxies with M$_* \gtsim 5 \times 10^{10}$ M$_{\odot}$ have up to an order of magnitude larger SSFRs, compared to  recent observations  by \citet{2009ApJ...697.1493S}.  To recover these observations we apply an empirical star formation efficiency in galaxies that scales with the host halo velocity dispersion as $\propto 1/\sigma^3$ during galaxy mergers. We find that this modification needs to be of stochastic nature to reproduce the observations, i.e. only applied during mergers and not during accretion driven star formation phases. Our choice of star formation efficiency during mergers allows us to capture both, the boost in star formation at low masses and the quenching at high masses, and at the same time produce a constant SSFR-stellar mass relation at $z\gtsim 4$ under the assumption that most of the observed galaxies are in a merger triggered star formation phase. Our results suggest that observed high-z low mass galaxies with high SSFRs are likely to be frequently interacting systems, which experienced bursts in their star formation rate and efficiency (mode 1), in contrast to low redshift  $z \ltsim 3$ galaxies which are cold accretion-regulated star forming systems with lower star formation efficiencies (mode 2).  

\end{abstract}

\begin{keywords}
galaxies: formation -- galaxies: high-redshift --  galaxies: interactions -- galaxies: starburst
\end{keywords}

\section{Introduction}
The cosmic star formation rate in the universe is shaped by the physical conditions that govern the growth of galaxies. An initial, probably 
short-lived, $t \sim 10^6$ yr,  episode of population III-dominated star formation at $z \sim 15$ \citep[e.g.][]{2010MNRAS.tmp..905M} is followed by a steady increase in cosmic population II/I star formation activity that reaches a peak around $z \sim 2$ \citep{2006ApJ...651..142H}, coinciding with a peak in the merger activity of galaxies \citep[e.g.][]{2001ApJ...561..517K,2008MNRAS.386..909C,2009MNRAS.397..506K}. The steady decline at lower redshifts is quite strong, however, allowing for  $\sim 50 \%$ of the present-day stellar mass to have been formed between $z= 0$ and $1$ \citep{2008ApJ...675..234P}.

Within the hierarchical $\Lambda$CDM framework, the initial increase in the cosmic  star formation is closely linked to the growth of the hosting dark matter haloes. The short cooling time-scale of halo gas favours efficient fuelling of star formation in galaxies, in particular at high redshifts and in dark matter haloes with $M_{vir} \ltsim 10^{12}$ M$_{\odot}$ \citep[e.g.][]{1977ApJ...215..483B,1977ApJ...211..638S,1977MNRAS.179..541R,2003MNRAS.345..349B}. More massive haloes at $z > 2$ however, reveal efficient feeding of their central galaxies as well. Cold streams of gas that enter the virial radius along cosmic filaments penetrate through the diffuse hot halo gas and reach the potential minimum of the halo \citep{2005MNRAS.363....2K,2008MNRAS.390.1326O,2009Natur.457..451D}. Massive galaxies at $z \sim 2$ show 
gas-to-stellar mass ratios of order unity suggesting that cold flows could indeed be a very efficient means of providing gas to galaxies \citep{2010ApJ...713..686D,2010Natur.463..781T}. 

At $z<2$, cold streams in massive haloes cease to exist and feedback effects control the cosmic star formation \citep[e.g.][see however Bouche et al. 2009]{2010MNRAS.402.1536S}. Different feedback sources are also held responsible for shaping the mass function \citep[e.g.][]{1986ApJ...303...39D,2003ApJ...599...38B,2006MNRAS.365...11C,2007ApJ...668L.115K,2010MNRAS.tmp..860O} or the colours of galaxies 
\citep[e.g.][]{2005ApJ...631...21K,2006MNRAS.370.1651C,2006MNRAS.368....2D,2008ApJ...680...54K}. 

The importance of cold accretion for the star formation rate of galaxies has  recently been the subject of various semi-analytic studies, which show that the evolution of the specific star formation rate (SSFR) of galaxies can be reproduced at $z \ltsim 4$ \citep{2009ApJ...700L..21K,2009arXiv0912.1858B,2010MNRAS.405.1690D,2010ApJ...713.1301K}. However, these models fail to reproduce the  observed SSFR of galaxies at $z > 4$ \citep{2009arXiv0912.1858B}, where in particular the SSFR is approximately  constant \citep{2009ApJ...697.1493S,2010ApJ...713..115G}.
One option is that a second mode of star formation enters at high redshift, perhaps driven by mergers \citep{2010arXiv1003.5180G,2010ApJ...714L.118D}, in contrast to the canonical cold gaseous disk instability model that elegantly reproduces the Schmidt-Kennicutt star formation relation at $z\ltsim 2.$

In this Letter,  we address the question whether it is possible to reconcile observations of specific star formation rates at $z>4$ with simple models of accretion-driven star formation at high-z. We focus on the question how star formation needs to be modulated to accomplish this task and derive empirical scaling relations for the star formation efficiency in galaxies at high $z.$
 
\section{Model}
The basic modelling approach we follow is similar to \citet{2010MNRAS.tmp..750N} and \citet{2009arXiv0912.1858B}.  In this approach the evolution of a galaxy within a dark matter halo is followed by calculating the accretion rate of cold gas, the associated star formation rate and the effect from supernovae feedback for a grid of dark matter haloes with varying masses starting at a fiducial redshift.   We approximate the accretion rate of cold gas onto central galaxies by $\dot{M}_{acc}=f_{b} \dot{M}_{DM}$, with $f_b=0.165$ as the universal baryon fraction and $M_{DM}$ the dark matter halo mass \citep{2009Natur.457..451D}. To simplify matters even more, we approximate the dark matter growth rate by an analytic fit $ \dot{M}_{DM}(z)\approx 6.6 f_{b}^{-1} M_{12}^{1.15} (1+z)^{2.25} $ M$_{\odot}$ yr$^{-1}$ \citep{2006MNRAS.372..933N}. Here  M$_{12}$ is the dark matter halo mass in units of $10^{12}$ M$_{\odot}$. Once the gas accretion rate as a function of time is known, one can calculate the gas mass in the central galaxy of a given dark matter halo. We assume that the gas settles onto a rotationally-supported disc hosted by an isothermal halo using the model of \citet{1998MNRAS.295..319M}, and convert the gas into stars using a global Schmidt-Kennicutt law $\dot{M}_*= \alpha_{*} M_{gas}/t_{dyn}$ \citep{1998ApJ...498..541K}. The fiducial value for the star formation efficiency is set to $\alpha_{*}=0.02$  \citep{1998ApJ...498..541K} and the dynamical time of the disc, $t_{dyn}=0.1 R_{vir}/V_c$ \citep{1999MNRAS.303..188K}. Here $R_{vir}$ is the halo virial radius and $V_c$ its circular velocity. Besides models with constant $\alpha_*$  we also investigate models with more general efficiencies that depend on the halo velocity dispersion. The feedback effects from supernovae onto the cold gas reservoir are included using the approach in \citet{1999MNRAS.303..188K} with $\dot{M}_{SN}= \epsilon E_{SN} m_{IMF} \dot{M}_{*} /V_c^2$. The energy per supernova is $E_{SN}=10^{51}$ ergs, $m_{IMF}=1/150$ M$^{-1}_{\odot}$ is the number of type-II supernovae per solar mass stars formed assuming a Chabrier IMF, $\dot{M}_{*}$ the star formation rate and $\epsilon$ is the feedback efficiency for which we assume a fiducial value of $\epsilon = 0.2 $. The gas reheated by supernovae feedback contributes to the hot halo and we assume that it does not cool down by $z=4$. Numerical simulations suggest that most gas in galaxies indeed never reheats before joining it \citep{2009MNRAS.395..160K}. In the following we show our results for solving the differential equations for the evolution of the stellar mass $M_*$, the cold gas mass $\dot{M}_{gas}=\dot{M}_{acc}-\dot{M}_{*}-\dot{M}_{SN}$, and star formation rate $\dot{M}_*$, by explicitly following the dark matter growth $\dot{M}_{DM}$ as a function of time using the fitting formulae  for the average growth rate given above. We start at $z=20$ with a grid of dark matter haloes ranging from $10^7 - 10^{10}$ M$_{\odot}$  and set the initial conditions of the galaxies in the haloes to  $M_*=M_{gas}=0$. We show in the following the properties of individual central galaxies that live in dark matter haloes with such average growth histories. Thus the galaxy properties refer to galaxies with average growth histories. In the following we assume a standard cosmology with $(\Omega_m,\Omega_\Lambda,h,\sigma_8) = (0.3,0.7,0.7,0.8)$. 

\section{Results}
\begin{figure}
  \begin{center}
    \includegraphics[width=8cm,angle=0]{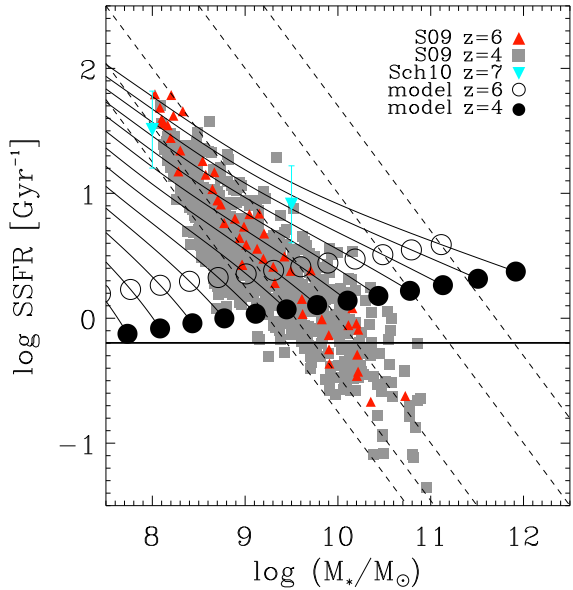} 
    \medskip
    \caption{The specific star formation rate as a function of stellar mass.  Each individual line shows the growth history of one central galaxy within a dark matter halo under the assumptions of the fiducial model. The initial halo masses at $z=20$ span a range from $10^7 - 10^{10}$ M$_{\odot}$. The solid black circles and open circles show the model SSFR at $z=4$ and $z=6$, respectively. The red triangles and blue squares show the observed data from \citet{2009ApJ...697.1493S} at these redshifts,  uncorrected for dust extinction and assuming an exponential decaying star formation rate with $\tau=100$ Myr. Typical errors are 0.3 dex. Light blue triangles show the $z=7 $ data from \citet{2010A&A...515A..73S}. Dashed lines show lines of constant star formation rate for $500$, $100$, $10$,  $3.5$ and $1.8$ M$_{\odot}/$yr. Latter two correspond to the reported completeness limit in \citet{2009ApJ...697.1493S} at $z=6$ and $4$, respectively. The \citet{2010A&A...515A..73S} sample is a combination of three different samples and we refer the reader to their paper on a discussion on the observational limits and incompleteness of their sample selection.  The solid horizontal line marks the inverse age of the universe at $z=4$.}  \label{fig1}
  \end{center}
\end{figure}

In Fig. \ref{fig1} we show the predicted SSFR of galaxies for the fiducial model and compare them to the observational data by \citet{2009ApJ...697.1493S}.    
  
Our results show the same trend as the non-accretion floor model of \citet{2009arXiv0912.1858B}. Applying an accretion floor of $M_{DM}=10^{11}$ M$_{\odot}$, below which no cold accretion occurs, gives good agreement with observed data at $ z<3$, but suppresses the formation of enough low mass galaxies at $z > 5$. For this reason we do not apply an accretion floor, and allow gas accretion in halos of all mass. The trend in SSFR is in stark contrast to the observed data. Most strikingly, low mass galaxies of $M_* \ltsim 5 \times 10^9 $ M$_{\odot}$ show orders of magnitudes too low star formation rates, while massive galaxies with $M_* \gtsim 5 \times 10^{10} $ M$_{\odot}$ show an order of magnitude too high star formation rates. Independently of these failures, the time-scales of the inverse SSFRs at $z=4$ are smaller than the age of the universe, indicating efficient early formation.   Fig. \ref{fig1} also highlights that the model SSFR-$M_*$ relation is significantly evolving from $z=4$ to $z=6$, indicated by open and filled circles, respectively. In contrast the observations shown in the same figure indicate almost no evolution.

 Before we continue comparing model results to observations it is worth pointing out potential observational biases.  The main source of uncertainty are the observed star formation rates. In Fig. \ref{fig1} we show SSFRs that are not corrected for dust extinction \citep{2009ApJ...697.1493S}. Including dust extinction results in higher star formation rates, shifting the SSFR-$M_*$ relation to larger values in SSFR, thus worsening the discrepancy between models and observations in particular at the low mass end, if these galaxies were to host significant amounts of dust. Another important source of uncertainty is the completeness of the observed samples. The  \citet{2009ApJ...697.1493S} sample is complete above $\log(SFR)= (0.25, 0.41, 0.54)$ for $z=(4,5,6)$. \citet{2009ApJ...697.1493S} argue that at low stellar masses their sample is not complete, and misses potentially low star forming galaxies. If indeed a large population of galaxies with star formation rates below these limits exists, the average derived SSFRs are over-estimates. However, independent of such corrections, the upper envelope of the SSFR-$M_*$  can be trusted, and poses particularly at low masses a challenge to models. 
   
\begin{figure} 
  \begin{center}
    \includegraphics[width=8cm,angle=0]{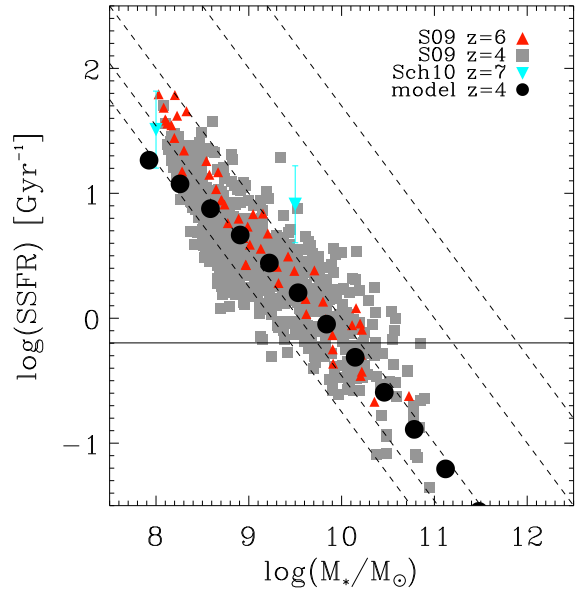} 
    \medskip
    \caption{Same as Fig. \ref{fig1} for a model with $\alpha_*(\sigma)=\alpha_0 \left(\sigma_0/ \sigma \right)^{\gamma}$ during mergers, with $\sigma_0=85$ km$/$s, $\gamma =3 $ and $\alpha_0=0.02$. During non-merger episodes $\alpha=\alpha_0$. We here assume that all galaxies undergo mergers at $z=4,5,6,7,8,9,10$. In reality galaxies will experience mergers at various different redshifts which will result in a scatter in the SSFRs. Thus the relation we are showing here represents the {\it maximum} SSFR-$M_*$ relation for individual galaxies. Galaxies not going through mergers will be lying closer to the evolutionary tracks shown in Fig. \ref{fig1}.  The $z=6$ model-relation looks virtually identical to the one at $z=4$, which is why we omit showing it in the figure.  }  \label{fig2}
  \end{center}
\end{figure}

We tested the behaviour of the SSFR for various star formation and feedback efficiencies. In general, increasing $\alpha_*$ leads to an increase in the stellar mass of galaxies within halos, and a lower SSFR. The latter is directly related to the scaling of the accretion rate with halo mass $\dot{M}_{acc} \propto M_{DM}^{1.15}$ and the lower reservoir of remaining gas in galaxies. The same trend is found for reducing the feedback efficiency $\epsilon$. SSFRs drop as galaxies form more easily and now live in less massive halos.   Our results suggest that a constant boost in star formation efficiency is not going to improve the low SSFR for galaxies with $M_* \ltsim 10^9$ M$_{\odot}$. What is needed is a boost in $\alpha_*$ that is not constant, but of a stochastic nature. Within the CDM framework, galaxy mergers are a natural candidate to provide such a behaviour. To recover the strong observed mass-dependence of the SSFR,  we propose a relation in which the star formation efficiency during mergers is a power-law of the halo velocity dispersion. We adopt the following relation during mergers:
\begin{equation}\label{eq1}
	\alpha_*= \alpha_0 \left(\frac{\sigma_0}{\sigma} \right)^{\gamma}
\end{equation}
with $\alpha_0=0.02$ as the fiducial value observed locally and $\sigma_0$  a characteristic velocity dispersion. The values of $\gamma$ and $\sigma_0$ are treated as  free parameters to obtain  the observed scaling in the relation between the SSFR and the stellar mass at $ z>4$. Our best-fit values matching the  $z=4$ \citet{2009ApJ...697.1493S} data   are $\sigma_0=85$ km$/$s and $\gamma =3 $. A similar scaling has been recently suggested in \citet{2009ApJ...700..262S} based on a porosity model for the inter-stellar medium. Within their model black hole feedback triggered by mergers plays a prominent role. The authors argue that black holes in low mass galaxies are  not massive enough to shut down star formation during their initial growth phase, but instead can possible help boost it \citep{2008MNRAS.389.1750A,2009ApJ...700..262S}, whereas massive black holes are able to prohibit star formation via AGN-feedback \citep[e.g.][]{1998A&A...331L...1S,2005ApJ...635L.121K}.  

\begin{figure}  \begin{center}
    \includegraphics[width=8cm,angle=0]{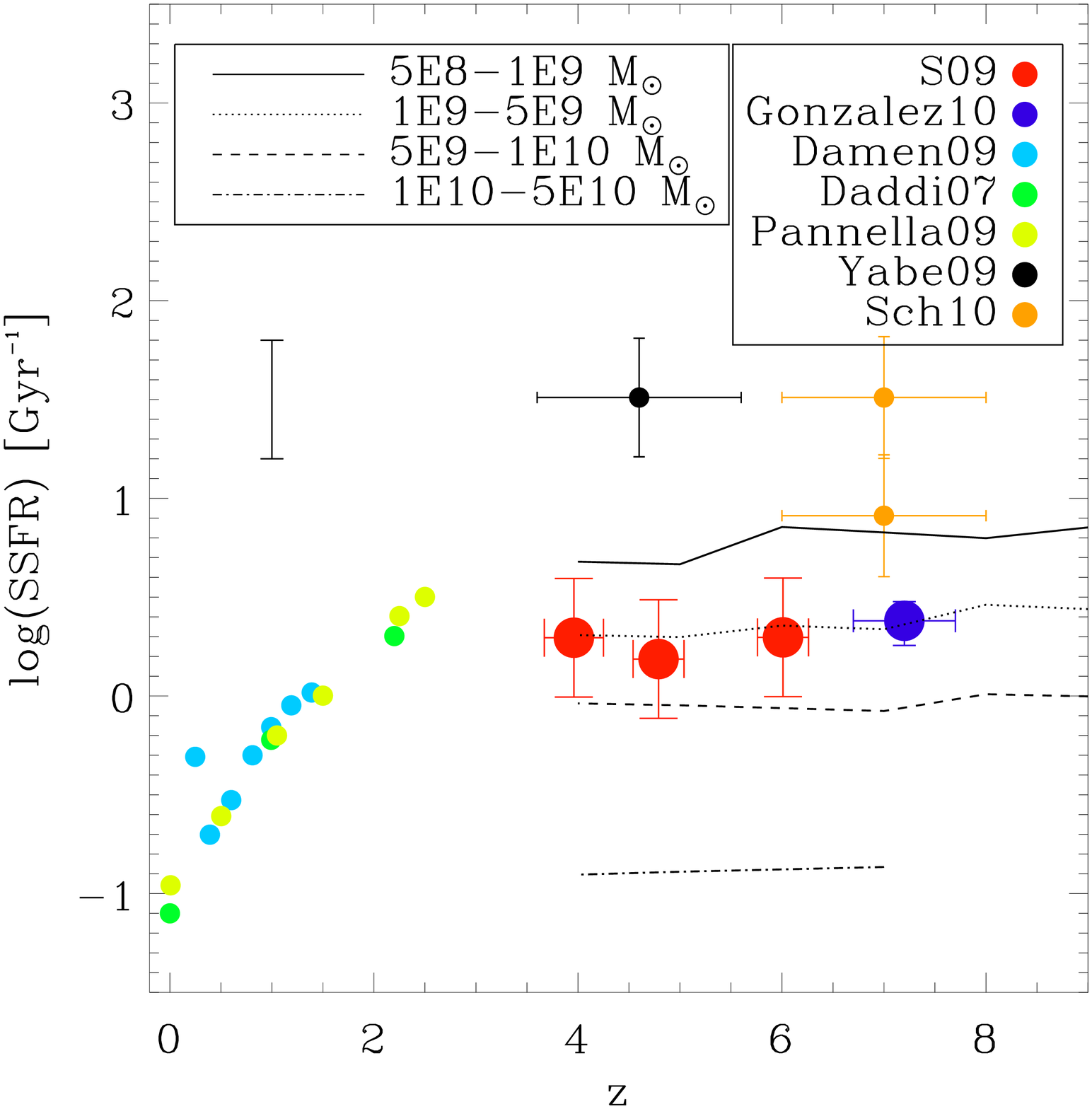} 
    \medskip
    \caption{The evolution of the SSFR for the merger triggered boosts in $\alpha_*$. For details on the low redshift data see \citet{2009arXiv0912.1858B}.  At $z<7$, typical errors in SSFRs are $ \sim 0.3$ dex \citep{2010ApJ...713..115G}, indicated by the error-bar on the top left. We show the average data for the \citet{2009ApJ...697.1493S} sample limiting ourselves to galaxies with $M_* \geq 10^9$ M$_{\odot}$ and star formation rates above the cited completeness limit. The \citet{2010ApJ...713..115G} data is the average for their sample, which has a medium mass of $M_* \sim 5 \times 10^9$ M$_{\odot}$. The lower and upper \citet{2010A&A...515A..73S} points refer to galaxies with $\log(M_*) \simeq 9.5$ and  $\log(M_*) \simeq 8$, respectively. The  \citet{2009ApJ...693..507Y} data includes galaxies with masses from $10^8$ to $10^{11}$ M$_{\odot}$ with a median value of $4.1 \times 10^9$ M$_{\odot}$.  We show model results for SSFRs averaged over mass bins  as lines. We here assume that the haloes within a mass bin contribute equally. }   \label{fig3}
  \end{center}
\end{figure}

We implement the stochastic nature of mergers in a very simplified manner to highlight their effects on the SSFR.  At redshifts $z=10, 9, 8, 7, 6, 5 $ and 4, we assume that every galaxy has an instantaneous merger. We assume that the burst is lasting 50 Myr with an efficiency given by Eq. \ref{eq1}. We furthermore neglect the stellar mass contribution of the merging partner to the remnant. Our model is clearly a simplification, and we tested the influence of our assumptions on the results  by letting mergers occur randomly during the evolution of galaxies, and changing the duration of the merger episode as well as adding stellar mass from a merger partner. It turns out that the exact frequency of mergers is not important, as long as there is enough time for gas accretion to happen after the merger. Given the high accretion rates at $ z > 4$ merger-triggered star formation can be frequent in our model, as long as the merger duration is not too long ($ < 100$ Myr). Adding mass from merger partners only shifts the relation slightly, which can be compensated for by adjusting $\sigma_0$.  Outside merger episodes, we apply the fiducial star formation  efficiency, i.e. $\alpha_*=\alpha_0$ and follow the evolution of our model galaxies. The evolution of the SSFR is presented in Fig. \ref{fig2}.  We find good agreement between the modelled SSFR-$M_*$ relation and the data of \citet{2009ApJ...697.1493S}. In addition we find that the our SSFR-$M_*$ relation does not change much with redshift, if we assume interacting galaxies are dominating the sample. As we pointed out above, the observed sample might be missing a population of low star forming galaxies. Such galaxies would be non-merging galaxies within our model, which show low SSFRs closer to the fiducial model shown in Fig. \ref{fig1}.   
To highlight the non-evolution of the SSFR we compare in Fig. \ref{fig3} our model predictions for individual galaxy masses with the observations in the literature. We find good agreement with the data of \citet{2009ApJ...697.1493S} and \citet{2010ApJ...713..115G} if we consider model galaxies between $10^9$ and $5 \times 10^{9}$ M$_{\odot}$. However, the data of \citet{2010A&A...515A..73S} and \citet{2009ApJ...693..507Y} show larger values in SSFR. Most of the discrepancy between these and the former data sets is due to different  star formation rate estimates that the authors assume, reflecting e.g. their different choice of dust extinction corrections and star formation histories. Interestingly, the latter surveys also predict no or only a week evolution from $z=4$ to $z=7$. Within our model we can achieve higher SSFRs to be in agreement with this data by adjusting our free parameter $\sigma_0$.  Our results suggest, that if indeed the sample of \citet{2009ApJ...697.1493S} is complete in terms of SSFR sensitivity, most if not all low mass high star forming galaxies have to be merger induced.

\section{Discussion and Conclusion}
In this Letter,  we present a study of the SSFR in $\Lambda$CDM models in which galactic star formation is driven by cold gas accretion onto a central galaxy, that parallels the growth of the hosting dark matter halo. Such models have been shown to reproduce the observed star formation rates of galaxies successfully at $z \ltsim 3$ \citep[e.g.][]{2009arXiv0912.1858B}, but fail at high redshifts  \citep[see however][]{2010arXiv1004.3545L}. 
The rise in SSFR  at $z\ltsim 2$ is due to the increased gas fraction. However, a new ingredient is required at higher $z.$ 
The two observational constraints of a high SSFR for low mass galaxies and non-evolution of the SSFR-stellar mass relation at $z>4$ cannot be met by a simple constant increase in the star formation efficiency $\alpha_*$ at $z >4$. However, an occasional boost in $\alpha_*$ triggered by mergers in between periods of low star formation efficiency can provide high SSFR. Galaxy mergers are known to trigger starbursts even in minor merger interactions ($M_1/M_2 < 10$, $M_1 \ge M_2$) \citep[e.g.][]{2010MNRAS.tmp..646P}. 

Observationally, merger-induced star formation at low redshifts $ z < 2$ is not the main driver of star formation \citep{2009ApJ...697.1971J}. This is consistent with the growth of dark matter haloes in simulations, which takes place through significant smooth accretion  events and not major mergers \citep[e.g.][]{2010arXiv1005.4058G}.  Our results indicate that the situation at $z > 4$ is rather different. Merging time-scale estimates using dynamical friction predict to first order a dependence on the mass ratios of the merging partners only, once expressed in units of the Hubble time \citep{2001MNRAS.328..726S,2008MNRAS.383...93B,2008ApJ...675.1095J}. An equal mass merger thus will take place approximately a factor of a few longer at $z=2$ than at $z=6$. This behaviour supports frequent interactions at  $z>4$ that can trigger star formation while at lower redshifts star formation takes mainly place during quiescent periods in between mergers.

While mergers certainly can boost star formation, AGN-regulated starbursts in mergers could provide a natural explanation for the non-evolution of the SSFR-stellar mass relation at $z>4$. If the effects of AGN-feedback are related to the dark matter potential \citep{2010MNRAS.405L...1B} a scaling of the star formation efficiency with the halo velocity dispersion can effectively produce a mildly or non-evolving  relation. We show that this is indeed the case for the empirical scaling that we adopt in this paper with $\alpha_* \propto \sigma^{-3}$ during mergers. A more physical model might appeal  to AGN growth and feedback which are much less efficient in low mass galaxies, thereby inhibiting quenching, while in massive galaxies, the central black holes are already fully grown and are responsible for quenching the SFR.

Enhanced merging at high redshift provides the SSFR boost that also lends itself to generating enhanced $\alpha/Fe$ in the resulting massive galaxies. We predict that there should be rare  low mass $\alpha/Fe$ "refugees", perhaps companions of massive ETGs, that have avoided the final merging fate and gas blow-out, but nonetheless carry chemical traces of their enhanced SSFR history.

The role of AGN remains to be elucidated. Quenching of star formation is  commonly attributed to AGN. This may be the case for massive galaxies at high redshift. However AGN may also play a role in boosting star formation in the low mass systems where the SSFR is enhanced.  A future test of the role of AGN will be to examine the residuals of a sample with measured AGN accretion rates and star formation rates in order to see whether for example the Eddington ratio
correlates (boosting) or anti-correlates (quenching) with SSFR.

Our results suggest that the majority of galaxies with $M_* \ltsim 10^{9}$ M$_{\odot}$ at $z > 4$ are interacting systems. This prediction can be tested with future upcoming observational missions, and should provide a strong test on the early build-up of galaxies. In a follow-up paper, we will investigate the individual contribution from merger triggered star-bursts  and AGN with respect to accretion-driven star formation in the context of  a full semi-analytic model.
\\

The authors would like to thank Dan Stark for kindly providing the observational data in electronic form and useful comments on the draft. SK acknowledges support from the the Royal Society Joint Projects Grant JP0869822.



\label{lastpage}

\end{document}